\documentclass[paper]{geophysics}
\usepackage{mathrsfs,amsmath, multirow, xspace}
\usepackage{hyperref}
\usepackage{makecell}
\begin{document}

\title{3D electromagnetic modelling and inversion: A case for open-source}

\renewcommand{\thefootnote}{\fnsymbol{footnote}}

\ms{...} % manuscript number

\author{
Douglas W. Oldenburg\footnotemark[1],\footnotemark[2] Lindsey J. Heagy\footnotemark[2], Seogi Kang\footnotemark[2] and Rowan Cockett\footnotemark[2] \\[16pt]
% affiliations
{\normalfont \small
\footnotemark[1]
Corresponding author: doug@eos.ubc.ca \\
\footnotemark[2]
Geophysical Inversion Facility, University of British Columbia
} \\[16pt]
% keywords
{
\normalfont \small
}
}

\footer{3D EM: A case for open-source}
\lefthead{Oldenburg et al., 2018}
\righthead{3D EM: A case for open-source}

\maketitle

\begin{abstract}
Electromagnetics has an important role to play in solving the next generation of geoscience problems. These problems are multidisciplinary, complex, and require collaboration. This is especially true at the base scientific level where the underlying physical equations need to be solved, and data, associated with physical experiments, need to be inverted. In this paper, we present arguments for adopting an open-source methodology for geophysics and provide some background about open-source software for electromagnetics. Immediate benefits are the reduced time required to carry out research, being able to collaborate, having reproducible results, and being able to disseminate results quickly. To illustrate the use of an open-source methodology in electromagnetics, we present two challenges. The first is to simulate data from a time domain airborne system over a conductive plate buried in a more resistive earth. The second is to jointly invert airborne TDEM and FDEM data with ground TDEM. SimPEG, Simulation and Parameter Estimation in Geophysics, (https://simpeg.xyz) is used for the open-source software. The figures in this paper can be reproduced by downloading the Jupyter Notebooks we provide with this paper (https://github.com/simpeg-research/oldenburg-2018-AEM). Access to the source code allows the researcher to explore the simulations and inversions by changing model and inversion parameters, plot fields and fluxes to gain further insight about the EM phenomena, and solve a new research problem by using open-source software as a base. By providing results in a manner that allows others to reproduce, further explore, and even extend them, we hope to demonstrate that an open-source paradigm has the potential to enable more rapid progress of the geophysics community as a whole.
\end{abstract}

\renewcommand{\figdir}{./figures} % figure directory

%%% ===========================================================================
%%% SECTION 1.
%%% ===========================================================================
\section{Introduction}

The potential for using electromagnetic (EM) geophysics to help solve problems in the geosciences has long been realized but success has not always been achieved. There are two primary impediments. The first is algorithmic; electromagnetics is intrinsically a 3D phenomenon and hence numerical 3D simulations and inversions are needed. The forward modelling, or simulation, is a powerful tool for building an understanding of how the earth is stimulated by an EM source. By solving Maxwell's equations in 3D and visualizing the resultant fields and fluxes, we obtain essential insight regarding the potential for using EM to delineate structure and to identify which EM fields are most important in the survey design process. Forward modelling is also the crucial component of the inverse problem and it needs to be efficient because it must be carried out many times. The second impediment lies in the ever-more complicated multi-disciplinary geoscience questions that need to be answered. The quantity and diversity of the data available to address a geoscientific question has increased significantly over the last decade. As a result, the quality of interpretations has the potential to be greatly improved. Achieving this will require advancements in computational components, data integration methodologies, and our ability as researchers to communicate across disciplinary lines. In practice, these presupposes that we have mechanisms for integrating diverse data types.

\begin{figure}
    \begin{center}
    \includegraphics[width=0.8\columnwidth]{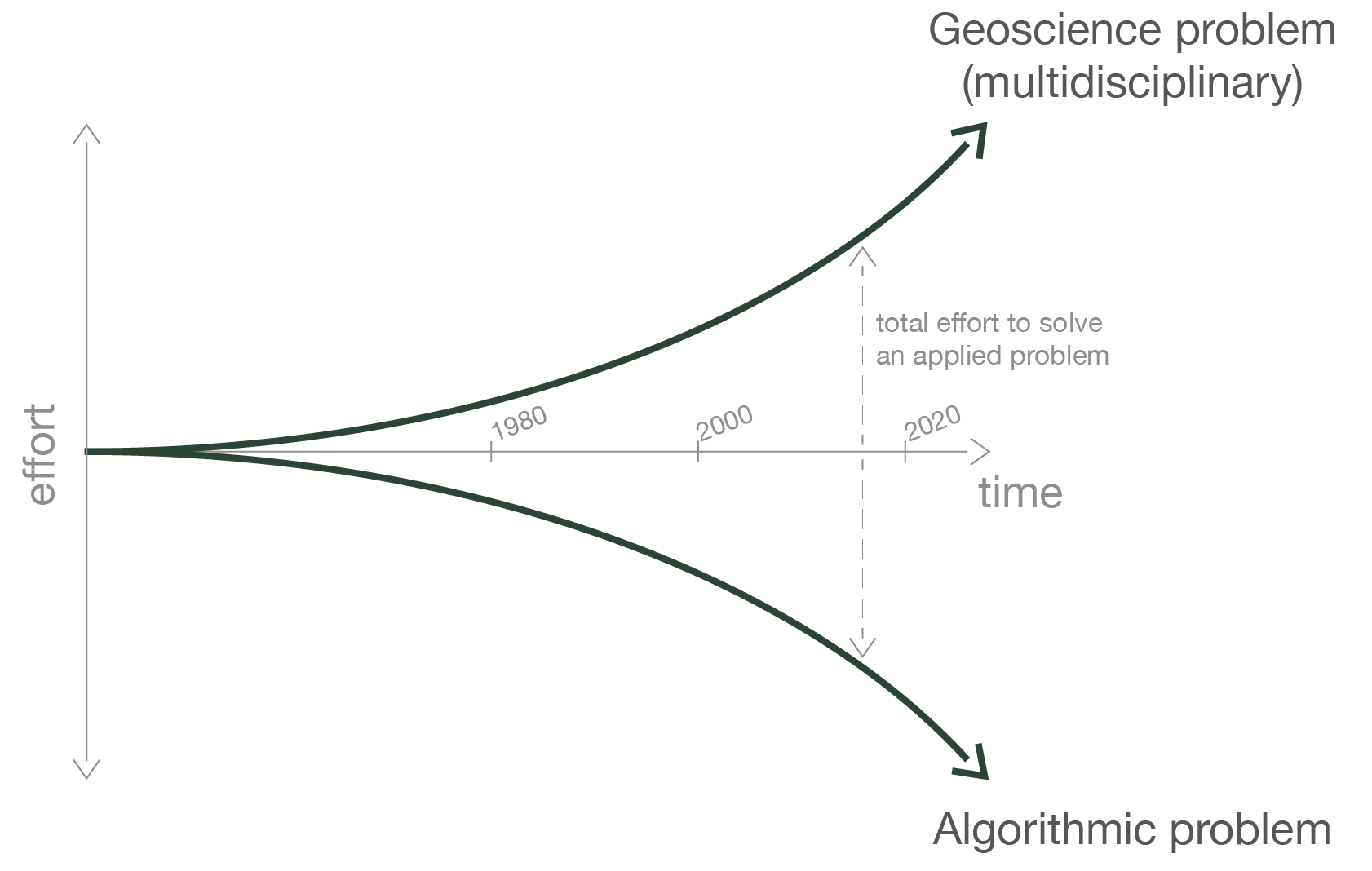}
    \end{center}
\caption{
    The plight of a researcher in geoscience. The horizontal axis reflects time, and the vertical distance away from this axis reflects the effort that a researcher must invest in order to solve a problem. The lower curve represents the effort required to solve an algorithmic problem and the upper curve represents the effort required to solve an integrated multi-disciplinary geoscience problem using algorithms. The distance between these two curves represents the total effort required to solve a state-of-the-art applied geoscience problem.
}
\label{fig:diverging-curves}
\end{figure}

These two impediments are intimately linked; more complicated geoscience problems require new developments of numerical tools, and the ability to obtain higher resolution images of the subsurface leads to further geoscience questions. This poses significant challenges for a researcher who wants to solve a modern geoscience problem using EM; this challenge is conceptualized in Figure \ref{fig:diverging-curves}.

The horizontal axis reflects time in calendar years. The lower solid line represents the effort required to solve an algorithmic problem. In 1980, for a graduate student to solve a problem at the state-of-the art, a typical algorithmic milestone might require that a 1D forward or inverse problem in electromagnetics be solved. Today the representative state-of-the art forward modelling might be a 3D time domain EM problem with dispersive physical properties and the inversion might incorporate thousands of transmitters. Although the field as a whole has advanced to a state where 3D forward modelling and inversion with thousands of sources is feasible, an individual graduate student who is interested in making an algorithmic contribution may have to start from scratch or from a code written by a previous graduate student (which in all too many cases, is essentially the same starting point).

In Figure \ref{fig:diverging-curves}, the solid line at the top conceptualizes the effort required to contribute to the solution of an applied geoscientific question. This requires the use of developed algorithms to invert data, as well as to integrate other types of data to make an informed interpretation. This effort has also increased with calendar years. The data sets are more extensive, with more transmitters and receivers covering large areas, higher in quality and more diverse with potentially multiple geochemical, geological, hydrological and geophysical data sets available at one site. This provides the opportunity for in-depth questions to be asked and detailed interpretations to be made. The breadth of knowledge that a researcher requires to address such a multidisciplinary problem is what drives the challenge and complexity along this line.

The vertical difference between the top and bottom curves is the total effort needed to solve a problem. Algorithms, such as EM inversions, are required to work with data and to make interpretations, while applications and the data integration opportunities they present are catalysts for  further development of algorithms. This total effort continues to increase and it is becoming impractical, or impossible, for one person to cross the gap by both developing software algorithms and applying them. For those who attempt this, it takes longer to complete their project, and/or their research is marginalized because time and funding are limited and may run out. This is detrimental to the researcher and to the development of geophysics as a whole. It is a paradigm that needs to be changed and working in a collaborative open-source framework is a key ingredient for achieving this.

To add context, we consider a geoscience problem connected with groundwater. The goal is to help characterize the subsurface by determining the electrical conductivity; in this regard extensive surveys are currently being flown. A typical survey might have an area of 10km $\times$ 10km and  involve thousands of transmitters with a goal of getting meter-resolution in the vertical dimension and 10’s of meter-resolution horizontally. In many settings 1D simulation and inversion is sufficient for answering the question-at-hand, but as we move to higher quality interpretations and more complex settings, 3D simulations and inversions will be an important tool. Yet, even carrying out the 3D forward simulation is challenging and involves many components including: (a) designing an appropriate mesh and discretizing Maxwell’s equations; (b) solving the resultant system of equations; and (c) computing simulated field data and evaluating their accuracy. The computations must be fast because the forward simulation must be carried out many times to solve the inverse problem. For instance about 20 million forward modellings may be needed to solve a time domain inverse problem involving 1000 transmitters, 50 time steps, and a Gauss-Newton methodology. Today, the milestone of inverting such large data sets has been achieved, primarily through the combination of advances including: (a) development of better numerical solvers for the discretized Maxwell’s equations involving a single source (e.g. direct solvers MUMPS \citep{Amestoy2001, Amestoy2006}, Pardiso \citep{DeConinck2016, Verbosio2017, Kourounis2018}); (b) use of (semi-)unstructured meshes to reduce the number of variables \citep{Haber2007}; (c) separating the forward modelling and inversion meshes \citep{Yang2014a, Haber2014b}, (d) computing and storing sensitivities or applying the operations of a sensitivity times a vector; and (e) access to multiple cores (locally or on a cloud) so that subproblems can be solved in parallel. New advances will require that all of these elements be brought together and built upon.

\begin{figure}
    \begin{center}
    \includegraphics[width=0.8\columnwidth]{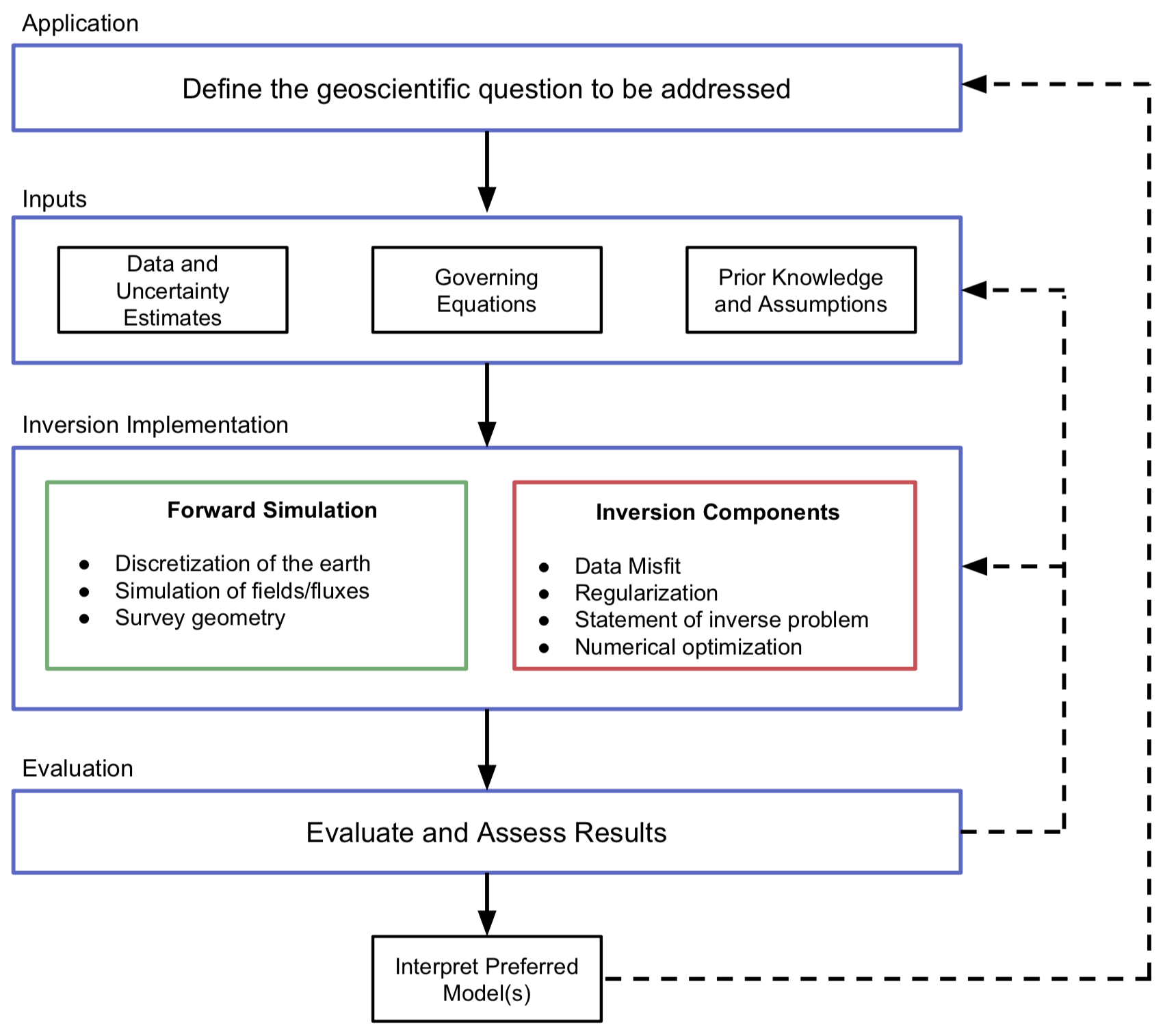}
    \end{center}
\caption{
    Components that must be assembled to use inversion to address a geoscientific question.
    The application motivates the data-collection strategy and necessary a-priori information.
    The inversion implementation includes forward simulation components such as the discretization
    as well as inversion components such as optimization.
    Obtaining a suitable model from the inversion is an iterative process
    that requires assumptions and choices (e.g. choice of the regularization functional) be tested and re-evaluated within the context of the
    initial geoscientific question.
    Adapted from \cite{Cockett2015}.
}
\label{fig:InversionWorkflowBullets}
\end{figure}

Unfortunately, there is a disconnect between the state-of-the-art and the state-of-what-the-researcher-has-access-to. In many cases, advancements have been made by a single researcher developing and implementing an algorithm. These algorithms may be available for others to use, but rarely is it the case that the code is sufficiently documented and well structured so that it can be refactored or built upon to address new questions. Thus, answering a new question requires that many of the components be re-implemented prior to making progress on the question at hand; the resulting duplication of efforts is inefficient and can slow progress. In addition, the increased level of complexity of the problem is reaching a stage where it is too much to expect a single researcher to solve. Consider for instance the 3D EM inversion problem. Some important component parts are illustrated in Figure \ref{fig:InversionWorkflowBullets}, With respect to the Inversion Implementation, there are several elements: meshing, discretization, numerical solvers, optimization routines, regularization functionals, simulation of partial differential equations etc., which must be composed. Each of these elements requires a specific expertise, which means a robust implementation will require input from researchers spanning several disciplines. All of the components must be assembled in order to solve the inverse problem, and the researcher who is performing the EM inversion needs to: (a) have access to all of these components; and (b) be able to communicate with appropriate specialists. The researcher also needs to interact with other geoscientists, including geologists and hydrologists, so that his / her inverse solution can help answer the question at hand.Lastly,  the research questions and algorithmic solutions are dynamic and ever-evolving. To make progress we need mechanisms for capturing algorithmic advancements in a modular, re-usable manner, as well as mechanisms for facilitating communication between researchers with expertise in each area.

\begin{figure}
    \begin{center}
    \includegraphics[width=0.55\columnwidth]{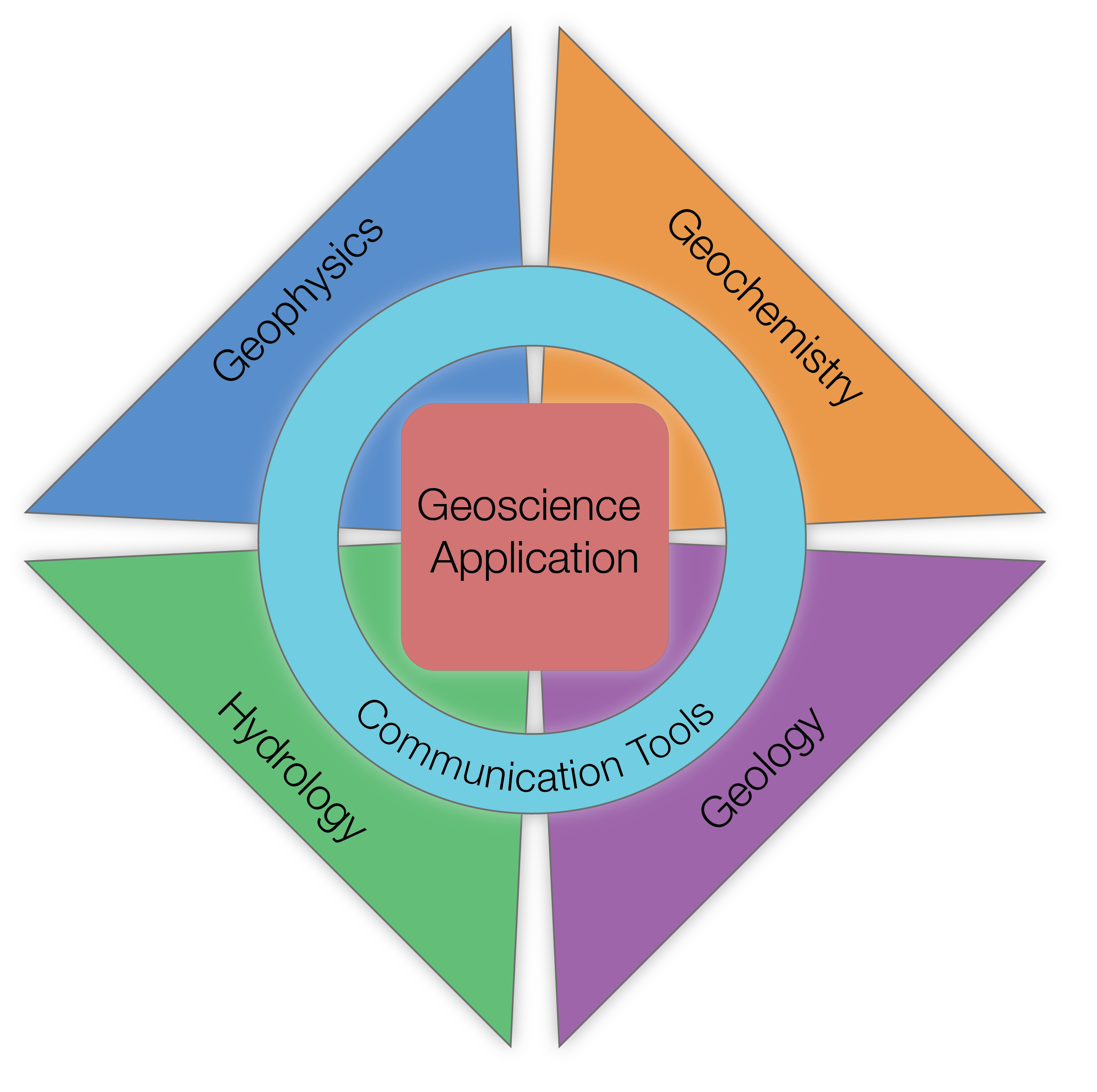}
    \end{center}
\caption{
    Solving a geoscientific problem requires knowledge and data from multiple disciplines. Enabling integration between disciplines requires communication tools which facilitate the transfer of data, algorithms and information across disciplinary lines.
}
\label{fig:multidisciplinary}
\end{figure}

There are a number of moving pieces that must be assembled to solve the algorithmic problem of estimating the electrical conductivity of the earth (the bottom curve in Figure \ref{fig:diverging-curves}). However, if we jump to the top curve in Figure \ref{fig:diverging-curves} and examine the interdisciplinary problem, electrical conductivity is not intrinsically of interest. The real value of EM is in integrating information about conductivity into issues connected with groundwater. Fundamentally these problems are in the realm of government or district water managers who want to know where the water exists, how much is available, how are the aquifers are being recharged (or not) etc. This requires knowing the geologic structure and hydraulic conductivity of the earth. EM geophysics certainly has a role to play, but integration with stakeholders or researchers from other disciplines, many of whom are not experts in geophysics and have different lexicons, is another level of challenge. Some of these challenges might be addressed by combining surveys from the different disciplines into process-based inversion (EM and fluid flow) and some addressed by understanding the information (often not geophysical) that is important, and incorporating it into the forward modelling and inversion algorithms. This level of integration requires that algorithms from different disciplines interoperate, and that researchers have resources to communicate across disciplinary lines (Figure \ref{fig:multidisciplinary}). The term “communication” in Figure \ref{fig:multidisciplinary} includes both qualitative communication of information, as well as quantitative communication of data and results.

We propose that an open-source paradigm can serve as a bridge between disciplines. With respect to the growing complexity of algorithms, open-source software provides additional opportunities to translate advances between fields. This can happen across disciplinary lines or across specialties within a discipline (e.g. potential fields and EM). With respect to the growing complexity of geoscience applications, both software and open-access training material enable data and results from one field to be incorporated into the methods and interpretations in another. In both cases, an open-source paradigm provides a space where researchers can connect and collaborate around a tangible set of tools and resources.

In this paper we focus on the algorithmic challenge of solving electromagnetic problems and provide context by forward modelling and inverting synthetic airborne EM data assuming an isotropic scalar conductivity function. However, the next generation of problems will require multiple airborne and ground EM surveys to provide information about the three physical properties ($\sigma$, $\mu$, and $\varepsilon$). Moreover, these properties can be frequency dependent, giving rise to induced polarization (IP) and viscous remanent magnetization (VRM)  in conductivity and magnetic permeability respectively. Thus there is an ever-growing need to solve such complex problems. In the next section we discuss some of the challenges faced by  researchers who will attempt to solve this next generation of problems.

\section{The researchers and their challenges}

Who will solve these problems and what are the challenges? Industry has potential to make progress in this regard but their research is generally confined to specific problems that provide financial benefits, usually in a relatively short time-frame. The task, therefore, largely falls into the hands of university researchers and research groups. Their work could be greatly accelerated if they had access to existing software that would allow them to explore solutions and build improvements and functionality into the software. Some of the most computationally efficient codes for handling large-scale data and carrying out forward simulation and inversion are in the hands of industry. However, using those codes to explore research questions is often not practical. Firstly, companies are often unwilling to share their codes because they may contain important technical advances and give the owners a competitive edge in a commercial market. Even if the codes are provided, they are ``fit-for-purpose''; they allow a particular type of data as input, have rigid control on inversion parameters, and provide a specific output. This makes sense for a company as it can make their workflows robust and more efficient. Research, on the other hand, is often more open-ended, and as such, it requires that the researcher have hooks into all aspects of the code, the ability to plug in new pieces, and the flexibility to experiment with methodology. Furthermore, many of the production-level codes are legacy codes which may or may not have associated testing, and are often optimized for performance making them impenetrable (especially older Fortran codes) to many researchers. Rather than make desired alterations it is often more efficient for a researcher to begin coding from scratch. As a result, graduate students and research groups around the world duplicate efforts and re-invent technology. This is impacting the current rate of research at universities and it will become a greater impediment as we attempt to solve more complicated multi-disciplinary problems. An open-source environment that facilitates collaborative development and communication is key to progress.

\subsection{Open-source development}

As the effort to solve a geoscience problem increases it is clear that solving the next generation of integrated geophysics problems requires effective collaborations between researchers with different backgrounds and skill sets. Other research communities, for example Astropy in astronomy \citep{Astropy2013}, Scikit-learn in machine learning \citep{Pedregosa2011}, and SciPy in numerical computing \citep{Jones2001}, have embraced the open-source approach for collaboration and cooperation on the development of software resources for research. Beyond simply making code publically available, scaling a collaborative effort beyond a handful of close colleagues requires the adoption of best practices such as version control, automated testing, comprehensive user and developer documentation, peer review of code, and issue tracking. Modern tools such as GitHub, Read the Docs, and Travis CI make this possible, accessible and scalable beyond close collaborators. One of the key factors in fostering large communities of contributors is the choice of a permissive license with clear terms, such as the MIT license (https://opensource.org/osd). Such licenses allow users to adapt the code for academic or industry-oriented purposes, and enables contributors from companies and universities to collaborate on a common set of tools.

The adoption of open-source practices for software development has the potential to greatly accelerate research in geophysics. Often new avenues can be explored with a few tweaks or combinations of existing pieces. When the building-blocks are open-source, they can be re-used, adapted, or assembled in new ways to explore a question, rather than having to start the implementation from scratch. For students and researchers exploring a problem unfamiliar to them, having access to source-code, which captures the details necessary for a successful implementation, can be an invaluable learning tool. In particular, non-linear inverse problems, such as those that we encounter in EM, require tuning parameters to be set and heuristics, such as a beta-cooling schedule, to be defined. These are critical aspects to the success of the algorithm, but are generally obfuscated in black-box codes. Within the open-source ecosystem, the Jupyter notebook is one tool that is enabling workflows and analyses to be widely communicated to other scientists and to the general public, thus making it easier to distribute results in a manner that is readily reproducible \citep{Perez2015}. These features speak to an individual getting up-and-running with software to address a research question.

Open-source practices facilitate and encourage peer-review and collaboration at the level of the implementation. The re-use and adaptation of code requires that software be documented, human-readable and developed in a modular, flexible manner. Peer-review of code before it is incorporated into the main code-base can reduce ambiguities, unnecessary complexity, as well as bugs in the implementation \citep{Wilson2014}.

In addition to well-established software packages such as ModEM and the Aarhus Workbench \citep{Kelbert2014, Auken2015}, and existing open-source packages such as MARE2DEM \citep{Key2011}, there is a growing ecosystem of open-source tools available for solving problems in electromagnetic geophysics . The open-source ecosystem also contains several other notable packages including empymod, fatiando, jInv, and pyGIMLi \citep{Werthmuller2017, Uieda2013, Ruthotto2017, Rucker2017}. These packages differ in objectives, capabilities, structure, interactivity, license, and coding language (commonly Python and Julia). Our efforts have been focused on the development of SimPEG \citep{Cockett2015, Heagy2017}.

SimPEG is an open-source framework and set of tools for simulation and gradient-based parameter estimation in geophysics. It includes finite volume simulations and inversion routines for a variety of geophysical applications including potential fields, vadose zone flow, DC resistivity, induced polarization, self-potential and electromagnetics \citep{Cockett2015, kang2016, Heagy2017, Miller2017, Cockett2018, Miller2018, Witter2018}. Simulations may be performed on several different mesh types, including cylindrically symmetric meshes, 3D tensor meshes and OcTree meshes. A staggered-grid, finite volume solution approach is used to solve the quasi-static Maxwell’s equations in both the time and frequency. The fields and fluxes, computed everywhere in the simulation domain, are readily accessible so that they can be easily visualized and explored. Such simulations and visualizations have proved valuable in the context of geoscience education \citep{Oldenburg2017} and can be a useful tool for understanding the physical processes that contribute to the data we observe.

\section{EM Simulations}
\begin{figure}
    \begin{center}
    \includegraphics[width=0.8\columnwidth]{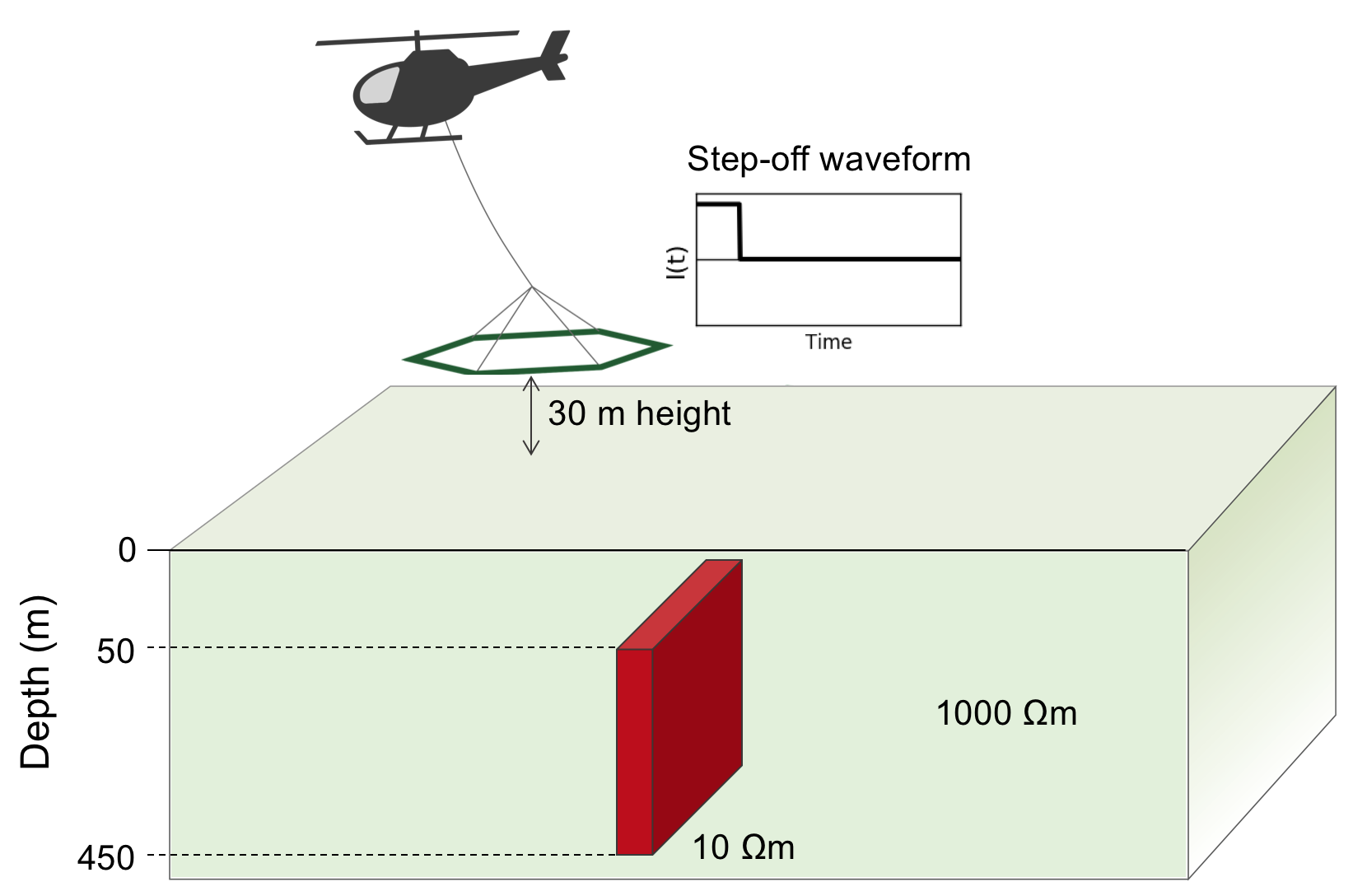}
    \end{center}
\caption{
    Airborne time domain EM survey over a conductive, vertical plate.
}
\label{fig:plate-setup}
\end{figure}

\begin{figure}
    \begin{center}
    \includegraphics[width=0.8\columnwidth]{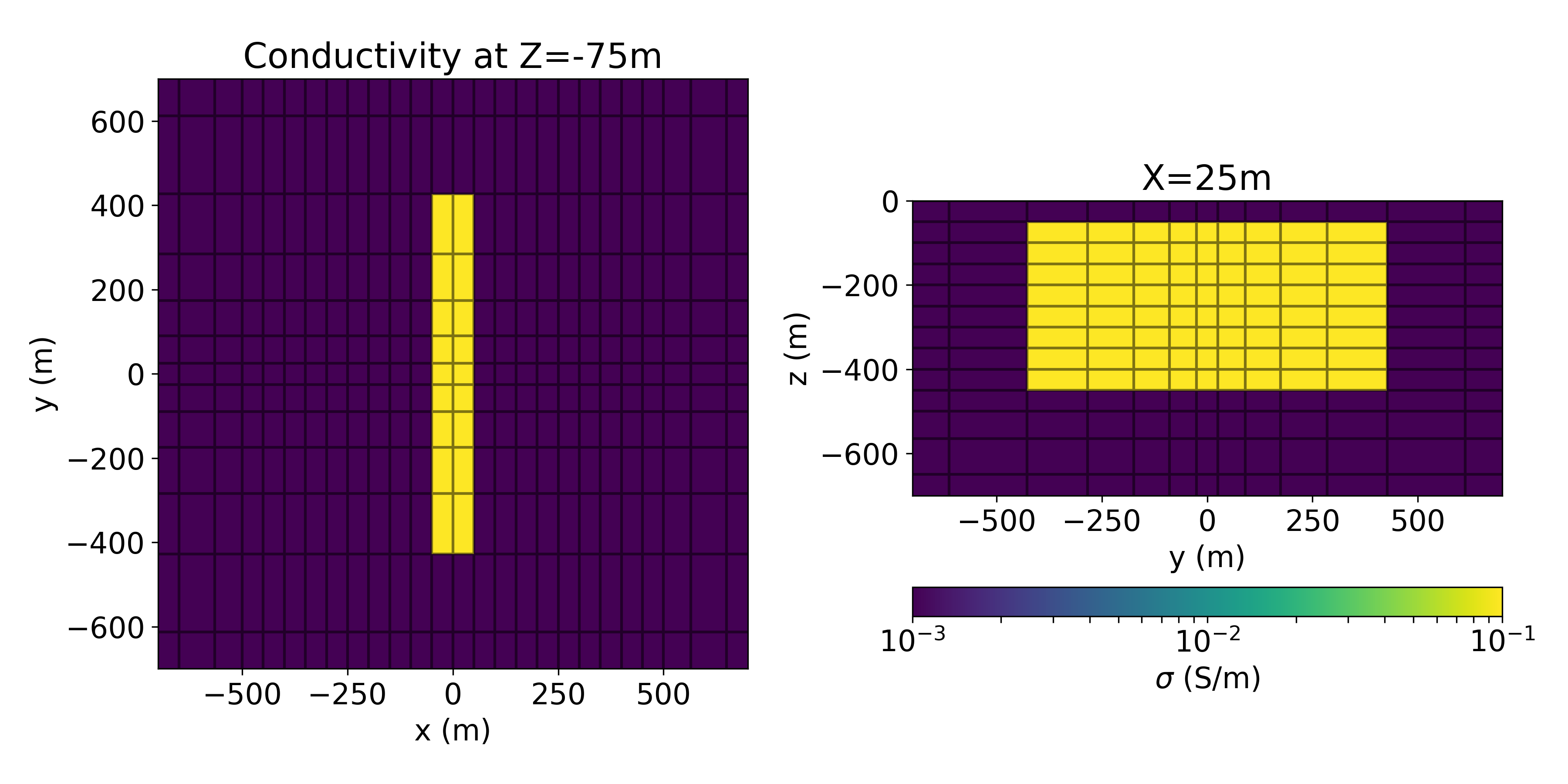}
    \end{center}
\caption{
    Depth slice (left) and cross section (right) through the model of a conductive plate (10 $\Omega$m) in a resistive half-space (1000 $\Omega$m).
    The mesh extends sufficiently far according to the diffusion distance for the time domain EM problem (see for example https://em.geosci.xyz).
}
\label{fig:plate-model}
\end{figure}

\begin{figure}
    \begin{center}
    \includegraphics[width=0.8\columnwidth]{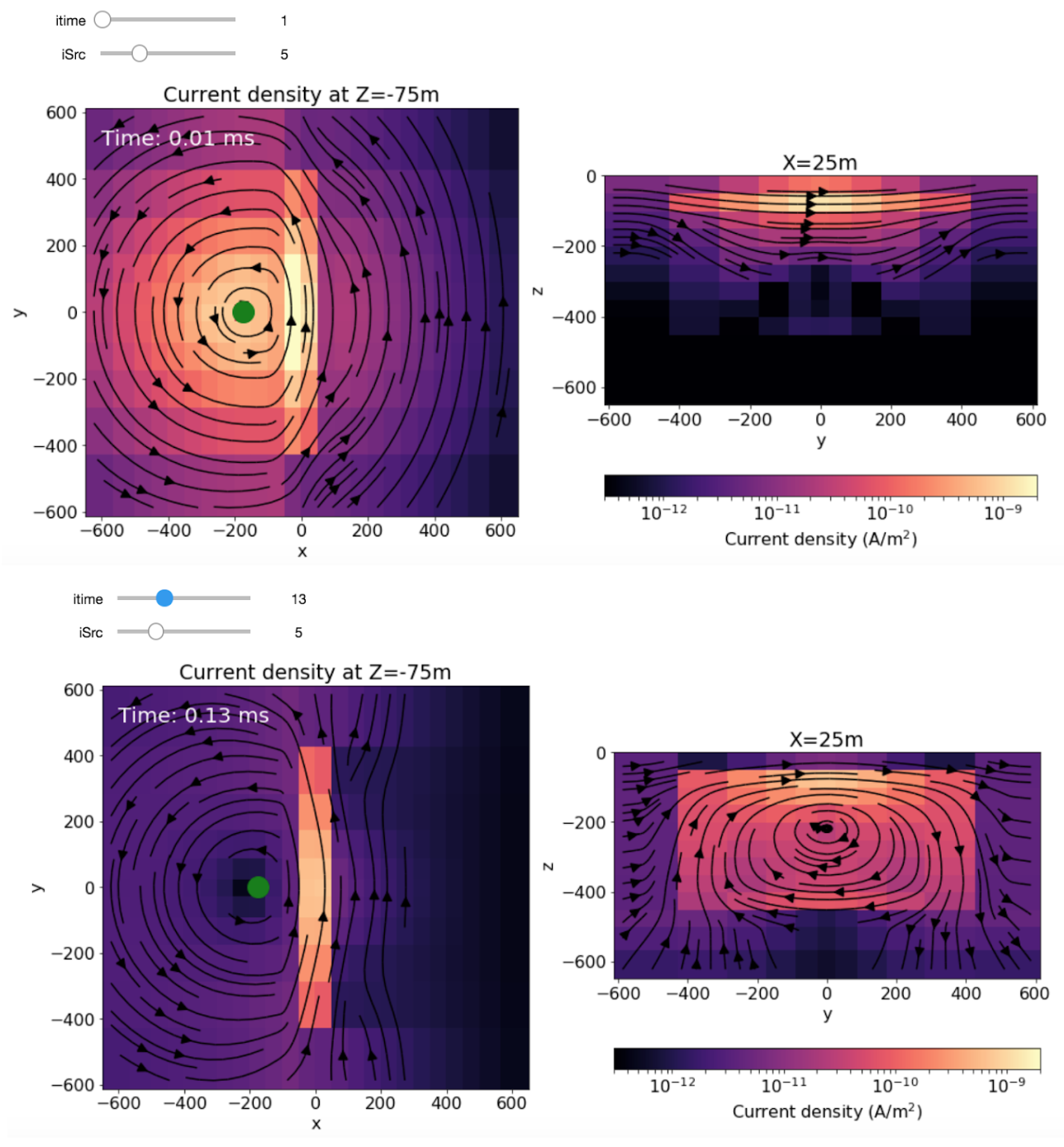}
    \end{center}
\caption{
    Snapshot of widgets in the Jupyter environment that a select a time and source location for which to view the currents. Here, we show the current density at 0.01 ms (top) and 0.13 ms (bottom) after shut-off. The panel on the left shows a depth slice and the panel on the right shows a cross-section. The green dot indicates the source location.
}
\label{fig:currents-widget}
\end{figure}

Many of the research questions we encounter start with an exploration of the governing physics. To demonstrate the value of this we choose a canonical model of a conductive plate in a resistive earth and perform an airborne TDEM simulation. The setup is shown in Figure \ref{fig:plate-setup}. The source loop has 10m radius and is located 30m above the surface; the source waveform is a step-off and the data are $db_z/dt$ measured at a receiver that is coaxial with the transmitter. The simulation is performed in 3D on a tensor mesh, which is shown along with the model in Figure \ref{fig:plate-model}; the mesh has 33600 cells and extends 2.8km in each direction to ensure that the fields have sufficiently decayed and satisfy the boundary conditions. A 100m thick conductive plate (10 $\Omega$m) is embedded 50 m below the surface and the background conductivity is 1000 $\Omega$m as shown in Figure \ref{fig:plate-model}. In addition to the data values, questions that often arise are: ``Where are the currents and how do they vary with time?'' and ``What is the relative contributions to the data from the inductive and galvanic currents?'' Insight for answering these questions, and understanding the EM induction process in general, can be obtained by carrying out a numerical simulation and viewing the currents and fields as a function of time. For example, in Figure \ref{fig:currents-widget} we show snapshots of the currents at two times. At early time (0.01 ms) the currents are primarily those that are channeled into the conductor (these are often referred to as galvanic currents). At later time (0.13 ms) induced vortex currents dominate. This transition between the two dominant current modes and their interactions can be further investigated as one scrolls through time using the slider widgets (https://github.com/simpeg-research/oldenburg-2018-aem, \cite{kang2018}).


\begin{figure}
    \begin{center}
    \includegraphics[width=\columnwidth]{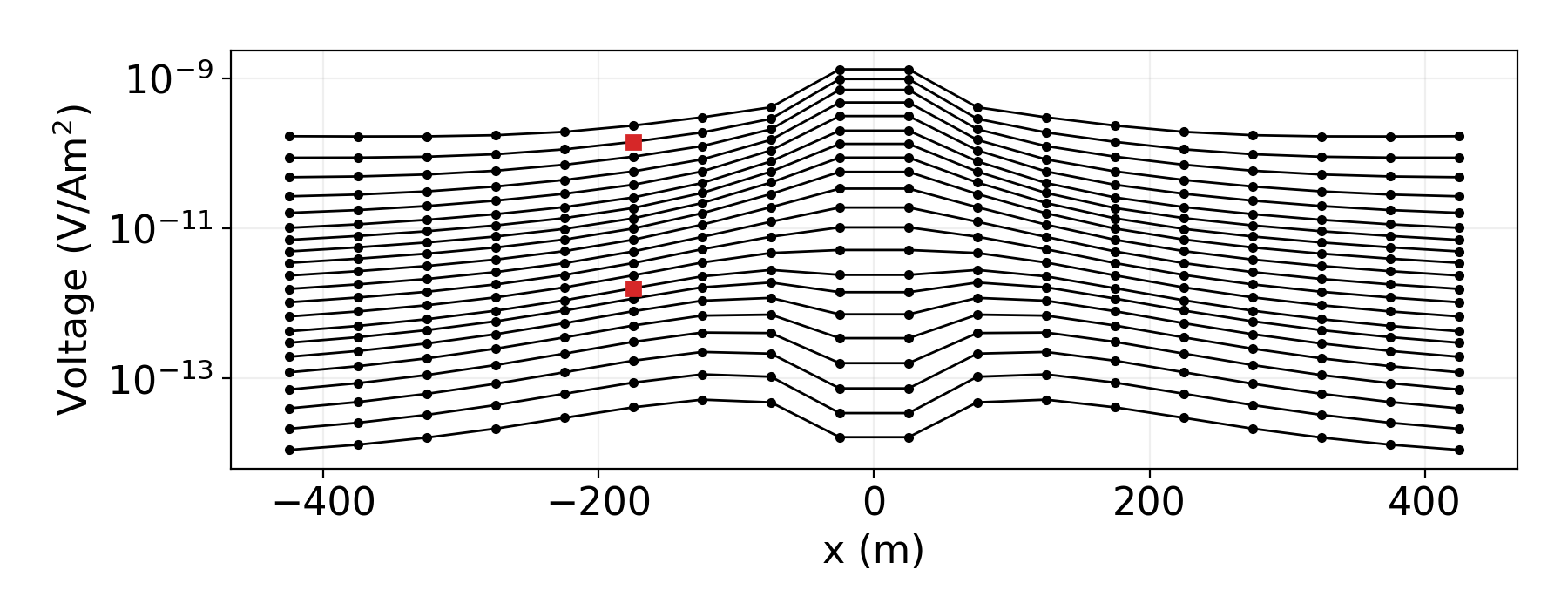}
    \end{center}
\caption{
    Simulated $db_z/dt$ data collected over the conductive plate. The data consist of 21 time-channels between 0.05 and 2.5ms. The red dots correspond to the times and source locations shown in Figure \ref{fig:currents-widget}.
}
\label{fig:aem-data}
\end{figure}

This interaction and the consequent time-varying fields are important factors in helping understand the data. Line plots of the simulated data are shown in Figure \ref{fig:aem-data}. The top two red dots correspond to the source location shown in Figure \ref{fig:currents-widget}. Offset from the plate, there is a transition between galvanic currents being the main contribution to the response at early times and vortex currents being the main contribution at later times. The transition between these behaviors and their connection to the data can be further understood by looking at the magnetic fields, which are shown in \cite{Heagy2018}.

\section{EM inversion}

Inversion is much more complicated than forward modelling; a framework and  workflow are required to extract meaningful information from EM data. Within SimPEG, we use a deterministic Tikhonov style approach, where an objective function consisting of a data misfit ($\phi_d$) and a regularization function ($\phi_m$) is minimized:
\begin{equation}
\begin{split}
    \min_{\mathbf{m}} \phi(\mathbf{m}) = \phi_{\rm d}(\mathbf{m}) + \beta \phi_{\rm m}(\mathbf{m}) \\
    \text{s.t.} ~\phi_{\rm d} \leq \phi_{\rm d}^*, \quad \mathbf{m}_l \leq \mathbf{m} \leq \mathbf{m}_u
\end{split}
\label{eq:inv-prob}
\end{equation}
The quantity $\beta$ is an adjustable constant that provides a relative weighting between the components of the final objective function. It is often referred to as the tradeoff or Tikhonov parameter. Within this framework there are many specific formulations that a researcher might like to explore. They pertain to definition of: the data (linear, log, some function of observed fields); the misfit function (least squares,$l_p$-norm); the model parameters (1D, 2D, 3D, linear, log, parametric); and the regularization function (norm, reference model, incorporation of a prior knowledge). In addition, there are many ways to solve the optimization problem, even within the spectrum of gradient-based methods, and there is always the issue of selecting an appropriate Tikhonov tradeoff parameter. In general, the researcher will want to carry out multiple inversions using different definitions of important elements. The open-source software should be tailored to allow this to happen as seamlessly as possible. We present two examples. In the first we invert the airborne data acquired over the vertical plate. In the second, we carry out a joint inversion of frequency and time domain data acquired in the air and on the ground.

\subsection{Inversion of airborne TDEM data}

\begin{figure}
    \begin{center}
    \includegraphics[width=0.55\columnwidth]{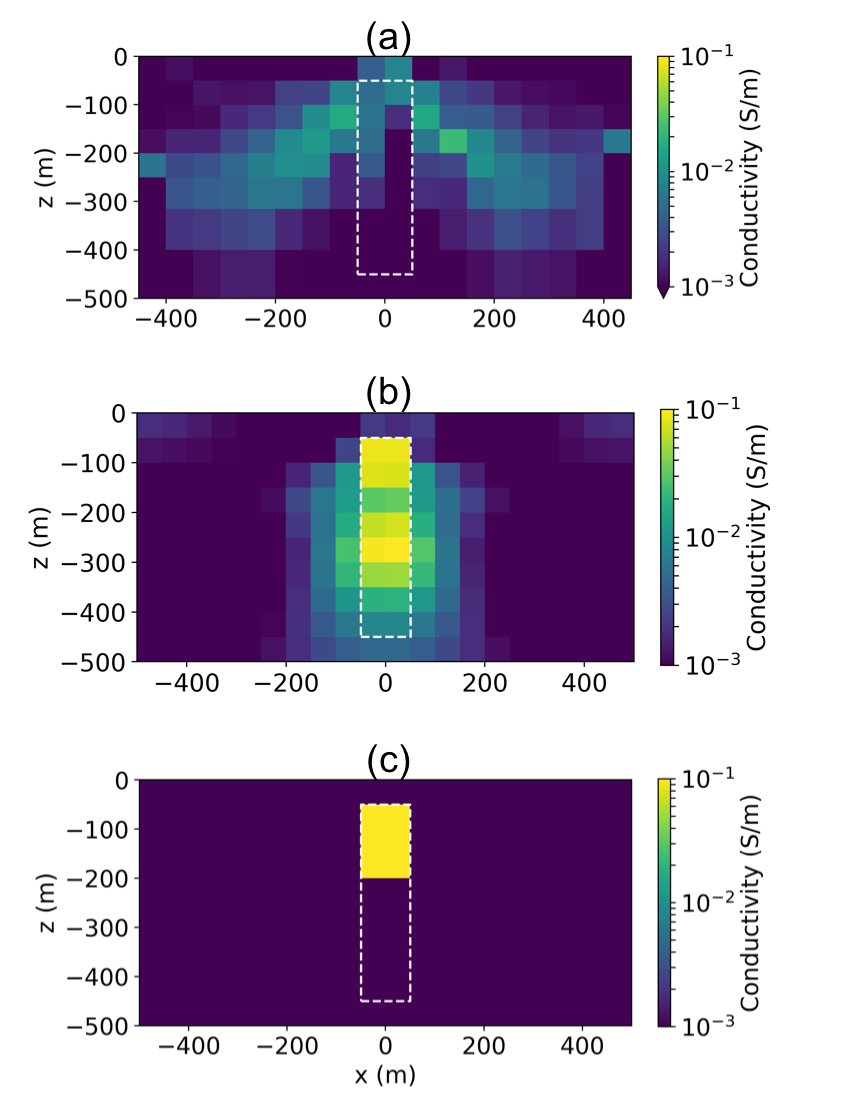}
    \end{center}
\caption{
    Models recovered by inverting the line of AEM data shown in Figure \ref{fig:aem-data}:
    (a) 1D inversion, (b) 2D voxel inversion, and (c) 2D parametric inversion. For more discussion
    on these inversions, see \cite{Heagy2018}.
}
\label{fig:aem-inversions}
\end{figure}

We now invert the airborne data shown in Figure \ref{fig:aem-data}. Although this is only a single line of data, we can use it to carry out multiple inversions that illustrate some of the flexibility of a modular framework. We begin with a  common approach for airborne EM inversions and invert the data in 1D. Each sounding is inverted independently using a cylindrically symmetric mesh for the forward modelling. The results are stitched together and the final image is shown in Figure \ref{fig:aem-inversions}a. The model has the typical ``pant-leg'' structure observed when the 1D assumption is imposed on a 2D or 3D model. Next, we perform a 2D voxel inversion. The simulation mesh is the same 3D tensor mesh shown in Figure \ref{fig:plate-model}, and the 2D model is projected along the line perpendicular to the flight line, thus the entire 3D volume for simulation is populated with values of electrical conductivity.  The recovered model is shown in \ref{fig:aem-inversions}b. Removing the 1D assumption and replacing it with  2D, which is much closer to reality, greatly improves the result. If it is known a-priori that the target is a compact body, then another option is to invert for a parametric representation of the model. Here, the inversion model is comprised of 6 parameters: the conductivity of the plate, the conductivity of the background, the center of the plate ($x_0$, $z_0$) and the height and width of the plate. Since the inversion is, in-principle, over-determined (there are more data than inversion parameters), no regularization is used \citep{Menke2012}. We use the same 3D forward simulation mesh as in the previous inversion; all that is changing is how we choose to represent the inversion model. The recovered model is shown in \ref{fig:aem-inversions}c. We recover good estimates of the top, width and electrical conductivity of the plate but to recover the depth-extent of the plate we may need to move to a 3D inversion. This can be accommodated in the SimPEG framework and for this example, would simply require changing the inversion model to a 3D representation of a plate. For further discussion on these inversions, see \cite{Heagy2018}.

\subsection{Joint inversion of airborne and ground data}

In this example, we carry out a joint inversion to recover electrical conductivity using data from different types of EM surveys. In previous years, software to invert data from each of the survey types has been developed, but merging the software to generate a joint inversion has been problematic. In a modular framework however, each datum can be considered to have its own forward modelling and sensitivity calculation so constructing an inversion that contains multiple surveys is straightforward.

\begin{figure}
    \begin{center}
    \includegraphics[width=0.8\columnwidth]{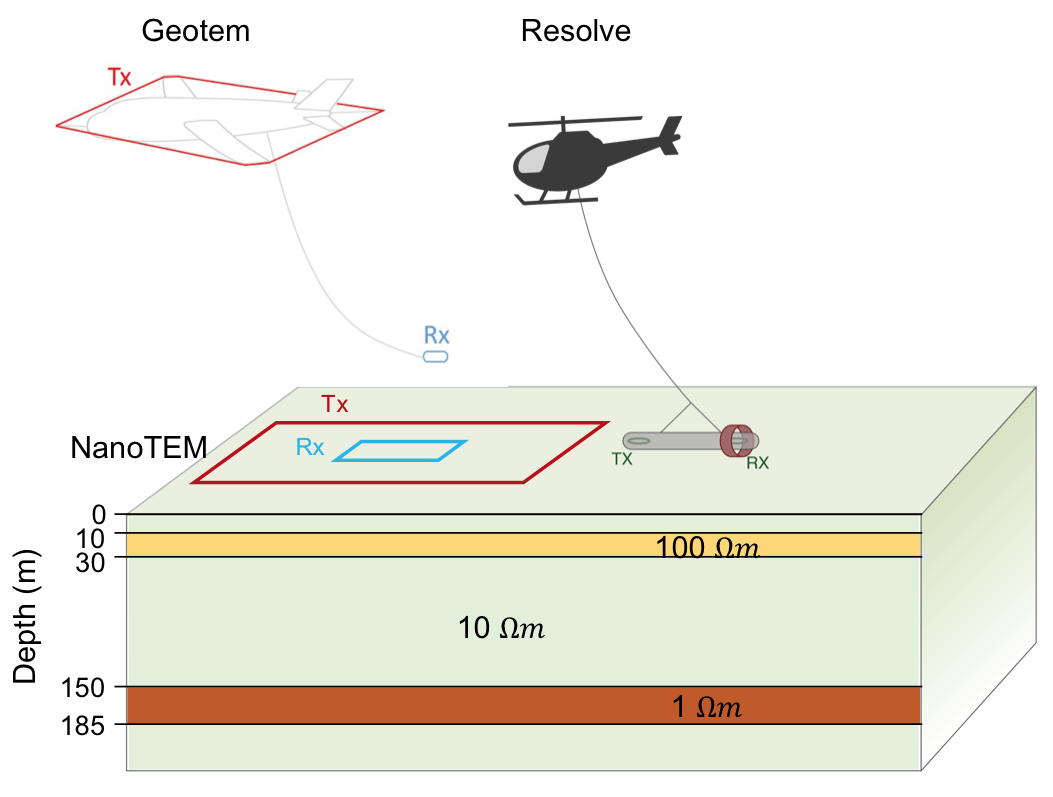}
    \end{center}
\caption{
    Three different EM systems: Resolve (airborne FDEM), Geotem (airborne TDEM), and NanoTEM (ground TDEM) over a layered-earth which includes a shallow resistor and a deep conductor.
}
\label{fig:joint-setup}
\end{figure}

For example, we consider a layered structure that includes a shallow resistor (100 $\Omega$m) and a deep conductor (1 $\Omega$m) embedded in half-space (10 $\Omega$m), as shown in Figure \ref{fig:joint-setup}. To image this conductivity structure, three different EM surveys are selected: Resolve (airborne FDEM), Geotem (airborne TDEM), and NanoTEM (ground TDEM), as illustrated in Figure 6. Each uses a loop transmitter and a receiver that measures $db_z/dt$. Each system has different sensitivity to the subsurface conductivity structure based upon its frequency-band and height above the surface. The frequency band of the Resolve system is 400Hz-130kHz, whereas the base frequency of the Geotem system is usually 30 (or 25 Hz). Even with very early time measurements (e.g. a couple microseconds), the Geotem system will have less sensitivity to the near surface compared to the Resolve system. However, for a deep conductor the Geotem system will have greater sensitivity due to its lower frequency band as compared to the Resolve system. Conversely, the ground NanoTEM loop, will show greater sensitivity to the near surface. Jointly inverting all three systems together should therefore be beneficial in resolving the layered conductivity structure.

Synthetic data sets are generated with the model shown in Figure \ref{fig:joint-setup} using  typical system specifications for Resolve, Geotem, and NanoTEM. Each of the three data sets is first inverted individually. Within the SimPEG framework, this is achieved by interchanging the survey parameters and PDE to be solved (e.g. FDEM or TDEM). The regularization, optimization, and other inversion elements are modular components which can be connected to each of the inversions. The initial model for each inversion is a 10 $\Omega$m half-space, and a 5\% percent standard deviation is assigned for the data-uncertainties. For the regularization, we employ a standard $l_2$-regularization,

\begin{equation}
\phi_{\rm m}(\mathbf{m}) = \frac{1}{2} \| \mathbf{W}_{\rm m} (\mathbf{m} - \mathbf{m}_{\rm ref}) \|_2^2
\label{eq:l2-regularization}
\end{equation}
where $\mathbf{W}_{\rm m}$ includes both a smallness term, which penalizes the difference between the model ($\mathbf{m}$) and the reference model ($\mathbf{m}_{\rm ref}$), and a first-order smoothness term, which penalizes changes in the model between neighboring cells.

\begin{figure}
    \begin{center}
    \includegraphics[width=0.8\columnwidth]{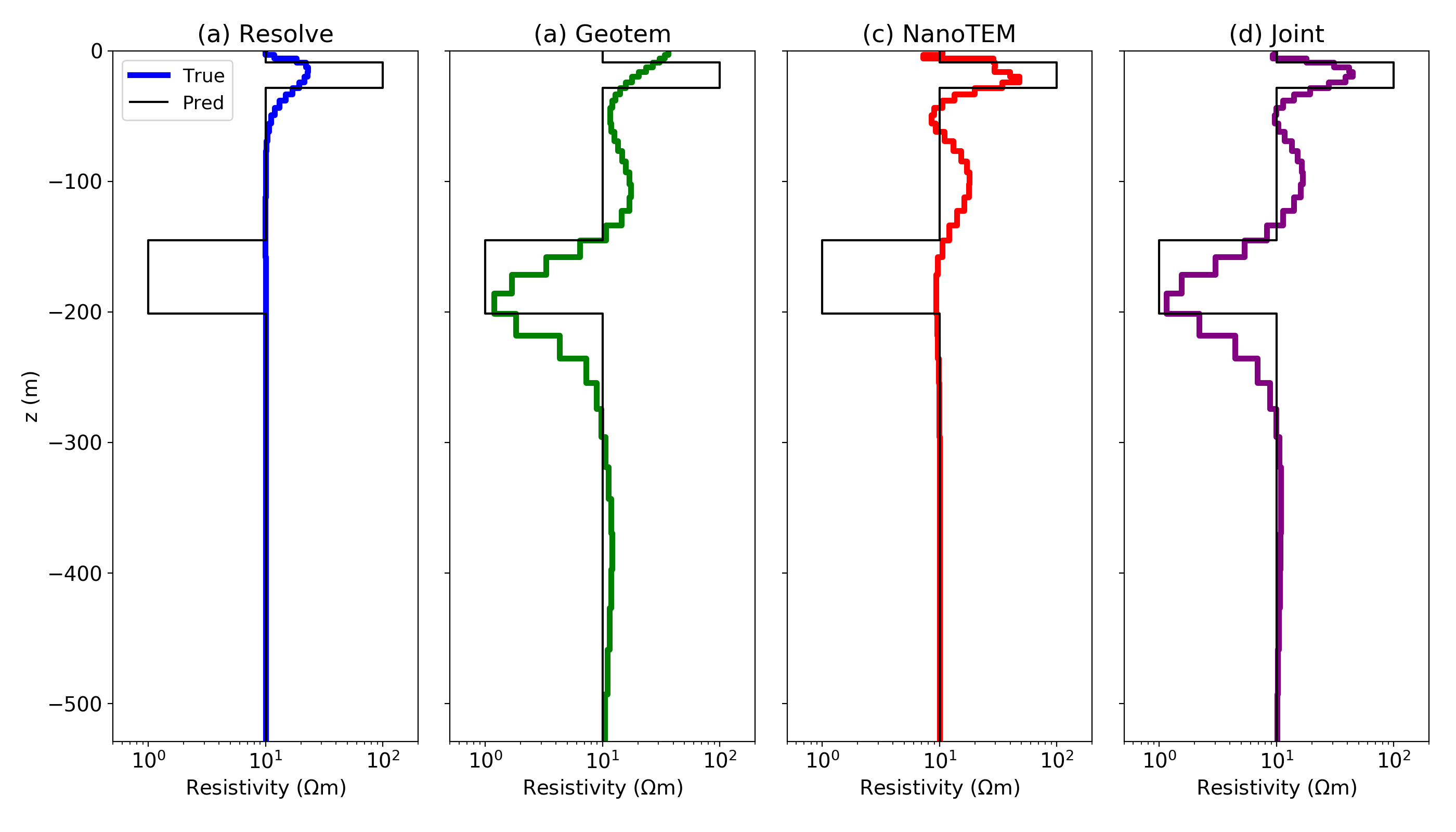}
    \end{center}
\caption{
    Models recovered by performing an inversion employing a smooth norm with: (a) the Resolve data (airborne FDEM), (b) the Geotem data (airborne TDEM), (c) the NanoTEM (ground TDEM) data and (d) jointly inverting all 3 data sets.
}
\label{fig:independent-inversions}
\end{figure}

\begin{figure}
    \begin{center}
    \includegraphics[width=\columnwidth]{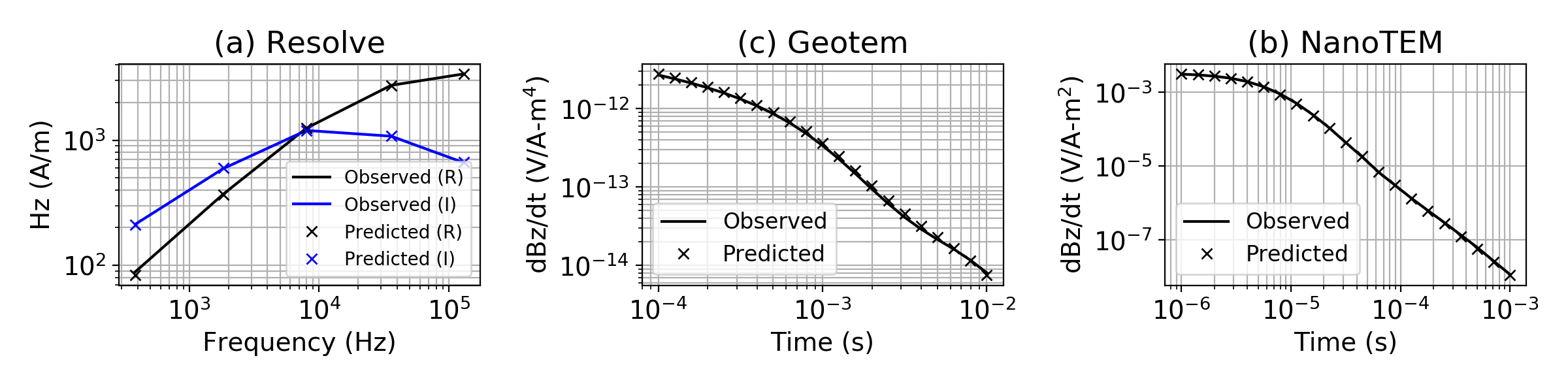}
    \end{center}
\caption{
    Comparison of observed and predicted data from each of the three inversions: (a) Resolve, (b) Geotem, and (c) NanoTEM. The joint inversion shows a similar level of data fit as those shown here. The black and blue colors distinguish real and imaginary values for the Resolve data. The solid line and cross marks indicate observed and predicted data, respectively.
}
\label{fig:obs-vs-pred}
\end{figure}

The recovered conductivity  models from the three separate inversions are shown in Figure \ref{fig:independent-inversions} (a, b, c); the associated data-fits are shown in Figure \ref{fig:obs-vs-pred}. Each recovers a shallow resistor but the value of the resistivity is underestimated compared to the true value. This is characteristic of all inductive source surveys. Of the three recovered models, the ground NanoTEM resistivity (red) shows the best match with the shallow resistor; this is due to its larger sensitivity  near the surface. Beneath the shallow resistor there are inversion artefacts in the NanoTEM result that are not seen in the Resolve inversion. Neither the NanoTEM nor the  Resolve inversions recover the deep conductor, however, it is well-recovered in the Geotem resistivity (green).  Although independently, none of these inversions reflects the earth model we aim to recover, each has complementary information indicating that a joint inversion may be a productive approach for recovering a representative earth-model.


Because all of the required forward-simulation machinery is implemented in one consistent framework, a joint inversion can be readily defined. Within SimPEG, a joint inversion is achieved by constructing a composite objective function to be minimized:
\begin{equation}
\phi = \underbrace{\phi_{\rm d}^{\text{Resolve}} + \phi_{\rm d}^{\text{Geotem}} + \phi_{\rm d}^{\text{NanoTEM}}}_{\phi_{\rm d}} + \beta \phi_{\rm m}
\label{eq:combo-objective}
\end{equation}
The ``global'' data-misfit, $\phi_d$ is composed of three terms, each with an independent forward modelling routine. These can be evaluated in parallel. A single regularization is used as we are only inverting for one model. In the code, the composite data misfit is simply constructed by adding the three, independent data-misfit terms together. Weighting terms, which scale the relative importance of each data misfit term can also be included, if desired. In the inversion results shown here, we weight the importance of each data misfit term equally. The inversion is terminated after each of the three data misfit reached a root-mean-square target misfit of unity.  The recovered model is shown in Figure \ref{fig:independent-inversions}d. The features recovered in the joint inversion are consistent with those in the independent inversions, with both the near-surface resistor and the deep conductor being imaged. Further exploration of the inversions, as well as reproducing the results shown here, can be achieved by using the Jupyter notebooks provided with this paper.

\begin{figure}
    \begin{center}
    \includegraphics[width=0.8\columnwidth]{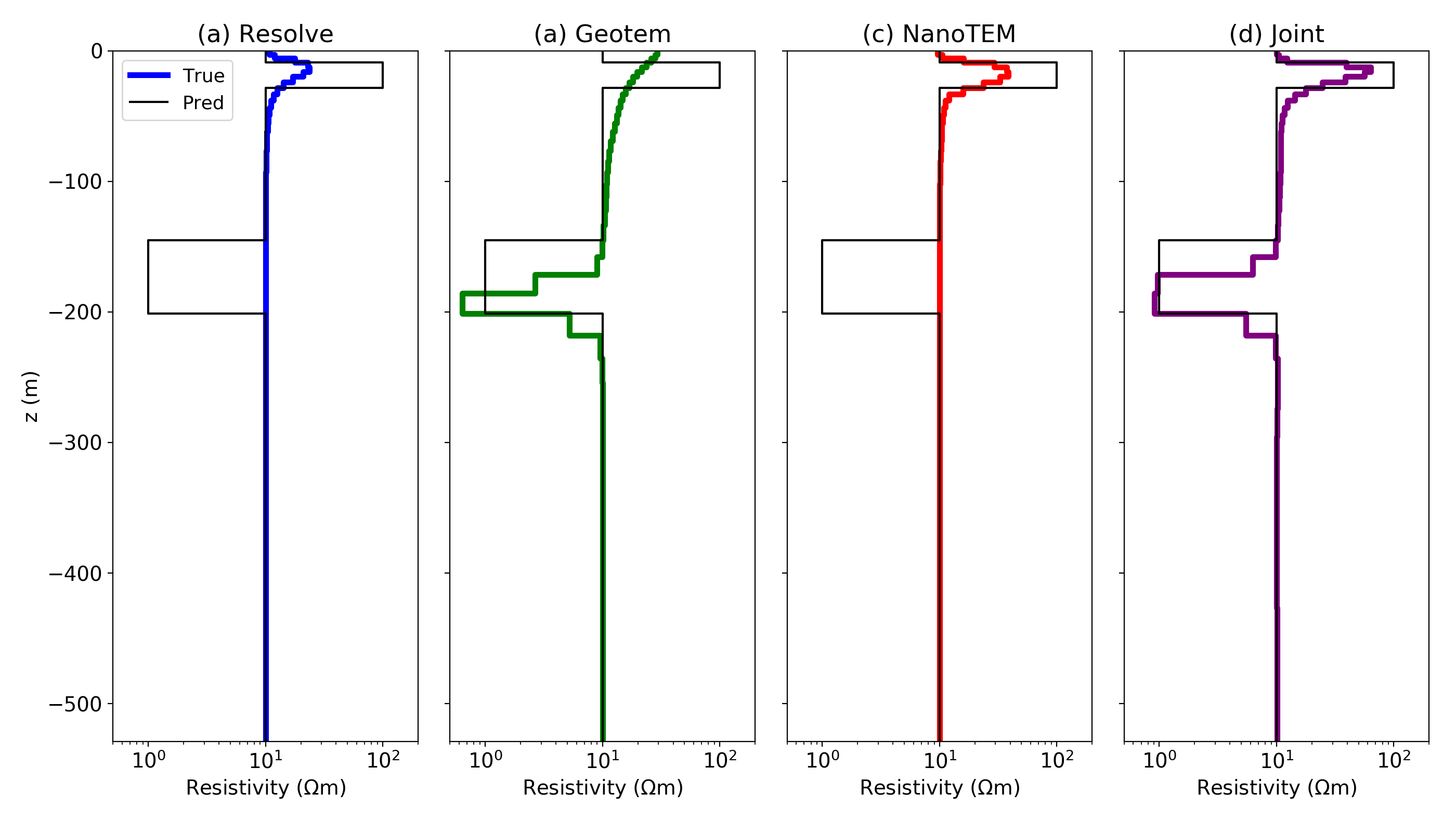}
    \end{center}
\caption{
    Models recovered by performing an inversion employing a blocky norm with:
    (a) the Resolve data (airborne FDEM), (b) the Geotem data (airborne TDEM), (c) the NanoTEM (ground TDEM) data and (d) jointly inverting all 3 data sets.
}
\label{fig:blocky-inversion}
\end{figure}

For all of the inversions shown in Figure \ref{fig:independent-inversions}, we used a standard $l_2$-norm for the regularization (equation \ref{eq:l2-regularization}). However, a model with different characteristics can be obtained by altering the regularization. For example, a blocky solution can be obtained by using the Lawson measure to approximate the $l_0$-norm \citep{Fournier2016, Lawson1961}. The development of sparse and blocky norms within SimPEG was originally conducted within the context of potential fields inversions. However, because  these developments are contributions to a modular framework, they can be readily applied to inversions using different physics for the forward simulation, including electromagnetics. In Figure \ref{fig:blocky-inversion}, an approximate $l_0$ norm is employed, and the Resolve, Geotem and NanoTEM data are independently inverted. The recovered models bare general similarity to those in Figure \ref{fig:independent-inversions} but inversion artefacts, seen previously in the NanoTEM inversion, have been suppressed. Panel (d) shows the joint inversion result using the blocky norm. The similarity between this result and that shown in Figure \ref{fig:independent-inversions}d provides confidence that the observed structures are supported by the data. Running several inversions with different norms can thus be a valuable approach for building an understanding of which features are primarily driven by the regularization and which are supported by the data. Having a computational framework that readily facilitates these changes reduces the researcher-overhead of having to learn new software or a new set of commands to perform each style of inversion.

\section{Discussion}

The examples we presented have demonstrated the use of the SimPEG framework for 3D forward modelling and 1D and 2D inversions of time-domain and frequency-domain EM data. Even for the simple model of a plate in a half-space, the physics is not necessarily intuitive, and 3D forward modelling is a powerful tool for building understanding. Recognizing that interrogating the physics is important across a range of problems, we have prioritized the ability to visualize and interrogate fields and fluxes through time or frequency. Interacting with the simulation results is facilitated by the availability of tools within the open-source ecosystem, in particular, the Jupyter framework and widgets, which enable easy construction of interactive ``research-apps''. With respect to the inverse problem, we demonstrated the power of a modular framework. In the first example, we transitioned between 2D voxel and 2D parametric inversions simply by changing the definition of the inversion model. In the second example, we experimented with different norms in the regularization functional and performed a joint inversion of multiple EM data sets.

These were synthetic examples in which the data and survey parameters are precisely known. The joint inversion is therefore straightforward to implement if the underlying framework is consistent and modular. In field situations however, there are complications that arise with respect to waveforms, system geometries, instrument gains, normalizations etc. and these must be supplied by contractors who acquired the data or from the instrument manufacturers. Again, open-source resources that make the information available or facilitate communication between diverse groups, in combination with a modular framework for including the information, greatly accelerates progress. Finally, although we have focussed on electromagnetic problems, the goal of jointly or cooperatively inverting multiple data types (e.g. potential fields and EM) is also readily assembled within a modular, open-source framework. This is not saying that the solutions are trivially achieved by connecting a few computational wires, but rather that the framework to explore solution-strategies exists. If experts who are working on these problems adopt an open-source paradigm and contribute their software and knowledge then, with time, these goals will be much more easily realized. The examples in this paper are a testimonial to this statement. Although each example was assembled by one or two individuals, the underlying infrastructure that enabled these examples to be run has been built by many more contributors across multiple institutions, and by researchers working in different specialties.

For example, the blocky-norm implementation we demonstrated was originally contributed by Dominique Fournier for use in potential fields problems. Because his contribution fits into the larger ecosystem, it was straightforward to use his work to run a blocky EM inversion. Another recent example is the refactoring of the OcTree mesh included in SimPEG. Joe Capriotti at the Colorado School of Mines re-wrote elements of the mesh storage and construction; this sped up the formation of the differential operators by 100$\times$. This advance can now be leveraged across all forward simulations implemented within SimPEG, and no changes are necessary in the forward simulation codes.

These collaborations are examples of researchers bringing their domain-expertise in one element of the framework and elevating the level of research that can be conducted by other researchers in the community. Naturally, as we broaden our perspective to include different disciplines within the geosciences,  there are more challenges to integration because of the differing lexicons and methodologies. However, through the continual development and refactoring of open-source building-blocks, we can make progress on reducing duplication of efforts and creating opportunities for methods, data and ideas to be transferred between researchers.

\section{Conclusion}

Unquestionably, the next generation of geoscience problems to be solved will be complex and multidisciplinary and, for electromagnetics, will likely involve multiple types of EM surveys coupled with process modelling and machine learning. Collaboration, access to previous research achievements and reproducibility will be key. In this paper we presented arguments for adopting an open-source model for geophysics and provided an introduction to some of the initiatives being pursued. An open-source paradigm encourages modular development and peer-review of software and of ideas. We demonstrated some of the implications of such a paradigm by discussing two examples that use SimPEG. In the first, we considered a 3D AEM forward simulation and demonstrated several approaches to parameterizing the inversion model, including: 1D, 2D voxel and 2D parametric. In the second, we performed a joint inversion of three different types of EM data. The components that were necessary to perform these tasks have been contributed to by a community of researchers, each bringing a unique perspective and expertise. With the use of Jupyter Notebooks, we illustrate how simulation and inversion research results can be shared and reproduced. Although our work has been contextualized within an EM framework, our comments on the pertinence of an open-source paradigm apply to all multidisciplinary geoscience problems and they are especially relevant for goals that involve jointly or cooperatively inverting data sets that have different underlying physics, such as those in potential fields, EM, seismology and hydrology. We see tremendous potential for accelerating geophysical research and for increasing the quality and reproducibility of research through the adoption of an open-source paradigm. We are not alone in our enthusiasm for this paradigm shift in geophysical research. Even within the realm of electromagnetic geophysics, there are many groups worldwide that have the same aspirations. Our challenge therefore, is not only to enable continued development within each group, but to promote opportunities for integration and collaboration between groups so that the open-source geophysical ecosystem grows coherently and sustainably.

\section{Acknowledgements}
We would like to thank the growing community around SimPEG who have contributed to discussions and improvements in the SimPEG code-base and community standards. In particular, thank you to Dominique Fournier, Thibaut Astic, Joe Capriotti, for their contributions to the electromagnetics, potential fields, meshing and regularization capabilities in SimPEG.

%% ==================================================================
\bibliographystyle{seg}  % style file is seg.bst
\bibliography{refs.bib}\end{document}